\documentclass{llncs}

\usepackage{amsmath}
\usepackage{amssymb}

\usepackage[T1]{fontenc}
\usepackage{graphicx}

\usepackage{amssymb}
\usepackage{adjustbox}

\usepackage{caption} 
\usepackage[T1]{fontenc}

\usepackage[pdfstartview=XYZ,
bookmarks=true,
colorlinks=true,
linkcolor=blue,
urlcolor=blue,
citecolor=blue,
pdftex,
bookmarks=true,
linktocpage=true, 
hyperindex=true
]{hyperref}

\usepackage{orcidlink}

\begin{document}

\title{Replay, Revise, and Refresh: Smartphone-based Refresher Training for Community Healthcare Workers in India}
\titlerunning{Smartphone-based Refresher Training for CHWs in India}

\author{Arka Majhi\inst{1}\orcidlink{0000-0002-5057-1878} \and
Aparajita Mondal\inst{2}\orcidlink{0000-0003-4609-2249} \and
Satish B. Agnihotri\inst{3}\orcidlink{0000-0002-0703-3185}}

\authorrunning{A. Majhi et al.}

\institute{
Indian Institute of Technology Bombay, Powai, Mumbai 400076, India
\email{arka.majhi@iitb.ac.in}\\
\url{https://arkadesignstudio.in}
\and
Tampere University, Kalevantie 4, Tampere 33100, Finland
\email{aparajita.mondal@tuni.fi}
\and
Indian Institute of Technology Bombay, Powai, Mumbai 400076, India
\email{sbagnihotri@iitb.ac.in}\\
}

\maketitle

\begin{abstract}
In India, community healthcare workers are the primary touchpoints between the state and the beneficiaries, such as pregnant mothers and children. Their healthcare knowledge directly impacts the quality of care they provide through home visits and community activities. Classroom in-person or traditional ways of training are found ineffective in imparting knowledge and render poor knowledge retention, which needs reinforcements through short, frequent revisions. Smartphone games on healthcare topics could be a promising solution as a refresher, as they can be scaled and tailored as per players' requirements. This study aims to check the differences in knowledge gain, pre and post-intervention, and, secondly, to check knowledge retention after six months. 270 CHWs or participants were recruited to evaluate different modes of refresher training and assigned into three equal groups of 90 each. The control group (CG) (n=90) was trained using the standard classroom method, which is usually followed. Intervention Group-1 (IG1)(n=90) was trained in a physical card game format, and Intervention Group-2 (IG2)(n=90) was trained in a smartphone game format. 4 sets of questionnaires were made by shuffling 45 questions based on immunization of equal weightage. The questionnaires were filled out by CHWs by hand and collected, evaluated, and analyzed. Paired t-tests were conducted to compare pre-post knowledge increments and repeated measure ANOVA to check for differences in knowledge retention. Results suggest a significant difference in scores in all three groups. A significant difference was observed between the physical and digital gameplay modes. Pre-post knowledge increment was higher in the digital mode (p<0.05), but knowledge retained was not significantly different (p=.4) in digital and physical card versions. Card games confirm their effectiveness in gaining knowledge when compared to classroom training. Through this research, we found that the gamified way of learning has an improved retention rate compared to the traditional training method.


\end{abstract}

\section{Introduction}

There are two main cadres of community healthcare workers (CHWs) with overlapping and sometimes complimentary job roles: Accredited Social Health Activists (ASHAs) and Anganwadi Workers (AWWs). The ASHAs bridge the gap between citizens and government by facilitating access to healtWhile services and receiving performance-based incentives \cite{Farah2015}. While the AWWs handle mother and child nutrition, early education, and overall development. They are forced to work in challenging conditions, with incomplete information, low compensation, and unrecognized or invisible efforts \cite{Verdezonto2021}. Their caregiving nature is motivated not only by their salary or incentive but also by the desire to earn respect, familiarity, and trust from the community \cite{Verdezonto2021}. Evaluation-based research highlights inadequate training and supervision as the primary challenges to the performance of CHWs \cite{Burger2022,Gopalan2012}.

According to the National Family Health Survey-5 (NFHS-5; 2019-20) \cite{NFHS5IndiaFactsheet}, immunization coverage for children aged 12-23 is 62.2\%, a marginal improvement from the last round, NFHS-4 60\% \cite{NFHS4IndiaFactsheet}, despite efforts from the government to provide free vaccinations through Universal Immunization Program (UIP) and mission Indradhanush. The Mother and Child Protection Card (MCPC) is given to pregnant women when they first register for pregnancy. It contains the child's immunization schedule and the necessary information for mothers. The CHWs' lack of sound knowledge of immunization, service schedules, and related information on MCPC reduces the effectiveness of their services \cite{Bag2017,Bashingwa2021}. Partial understanding of immunization schedules by CHWs often leads to partial child immunization in the community \cite{Kizhatil2019}. Hence, it becomes trivial for CHWs to understand the immunization schedule. Therefore, we chose the timeline or content of the child immunization schedule, which is derived from the Indian Academics of Paediatrician (IAP) guidelines \cite{Kasi2021} (referred in Table \ref{Tab:ChildImmunization}), Ante-Natal Care (ANC) or during pregnancy and Post-Natal Care (PNC) or after the birth of a baby, as the learning material.

Classroom in-person or traditional ways of training are found ineffective in imparting knowledge and render poor knowledge retention, which needs reinforcements through short, frequent revisions. The adoption of smartphones in India \cite{TRAI_Nov_23} is fast increasing. Smartphones are being given to the CHWs to maintain records of mothers and children. Smartphone-based games on healthcare topics could be a promising solution as a refresher training for the CHWs. Digital or smartphone-based games engage players through gameful interactions, are customizable according to the player's learning requirements, are playable by low-literate and less tech-savvy players, and reach a large audience by listing in app stores.

The study tries to answer the following research questions:
\begin{itemize}
    \item \textbf{RQ1:} Are games an effective alternative to in-person, traditional classroom refresher training?
    \item \textbf{RQ2:} Do physical and digital gameplay modes differ significantly in knowledge gains?
    \item \textbf{RQ2:} Do physical and digital gameplay modes differ significantly in long-term (after three weeks) knowledge retention?
\end{itemize}

\begin{table}[hbt!]
\centering
\begin{tabular}{|l|c|c|c|c|c|c|}
 \hline
 Vaccine name & Birth & \(1\frac{1}{2}\) months   & \(2\frac{1}{2}\) months & \(3\frac{1}{2}\) months  & 9 months & \(1\frac{1}{2}-2\) years\\
 
 \hline
 BCG & \checkmark &  &  &  &  & \\ \hline
 Hepatitis-B & \checkmark &  &  &  &  & \\ \hline
 OPV & \checkmark & \checkmark & \checkmark & \checkmark &  & \checkmark \\ \hline
 IPV &  & \checkmark &  & \checkmark &  & \\ \hline
 Pentavalent &  & \checkmark & \checkmark & \checkmark &  & \\ \hline
 PCV &  & \checkmark &  & \checkmark & \checkmark & \\ \hline
 Rota &  & \checkmark & \checkmark & \checkmark &  & \\ \hline
 MR &  &  &  &  & \checkmark & \checkmark\\ \hline
 JE &  &  &  &  & \checkmark & \checkmark\\ \hline
 DPT &  &  &  &  &  & \checkmark\\ \hline
 
 \end{tabular}
 \caption{Child Immunization schedule}
 \label{Tab:ChildImmunization}
 \end{table}

\subsection{Related Works}

Previous researchers conducted different methods of training CHWs. A group of researchers suggested using community radio to broadcast training materials to the masses \cite{Kumar2015,Kumar2015a,Kumar2013,Ward2020,Yadav2019,Yadav2019a,Yadav2017}. Researchers created and collected short informative videos on community healthcare and distributed them to CHWs through memory cards for mobile phones for offline viewing \cite{Javaid2017,Ramachandran2010}. After watching the short videos, researchers conducted quizzes through phone calls \cite{Shah2017} and quiz apps on their smartphones \cite{Majhi2021} to refresh their knowledge, followed by an incentive of talk time in the form of cell phone balance. Instructional illustrations were also tried to impart procedural knowledge to the CHWs \cite{Tulaskar2020}. Playful activities based on Augmented Reality were tested for collaborative play and refresher training on immunization \cite{Majhi2022}.

\section{Methodology}

\subsection{Physical and Digital Card Game}
A deck of 60 cards is designed to represent all the vaccines and services related to mother and child care. The cards are divided into four silos or foundations: Children below one year, Children above one year, Antenatal care and Postnatal care. The game aims to sequence the cards as per their schedule. Each player takes turns putting their card from hand to the foundations, towards completing the sequence forward or backward, starting from a point in each of the four foundations. The player emptying their hand first wins the game. A digital version of the game with the same rules was designed to compare the effectiveness of both modalities through user testing. The details of the game are not in the scope of this paper.

\begin{figure}
  \includegraphics[width=\textwidth]{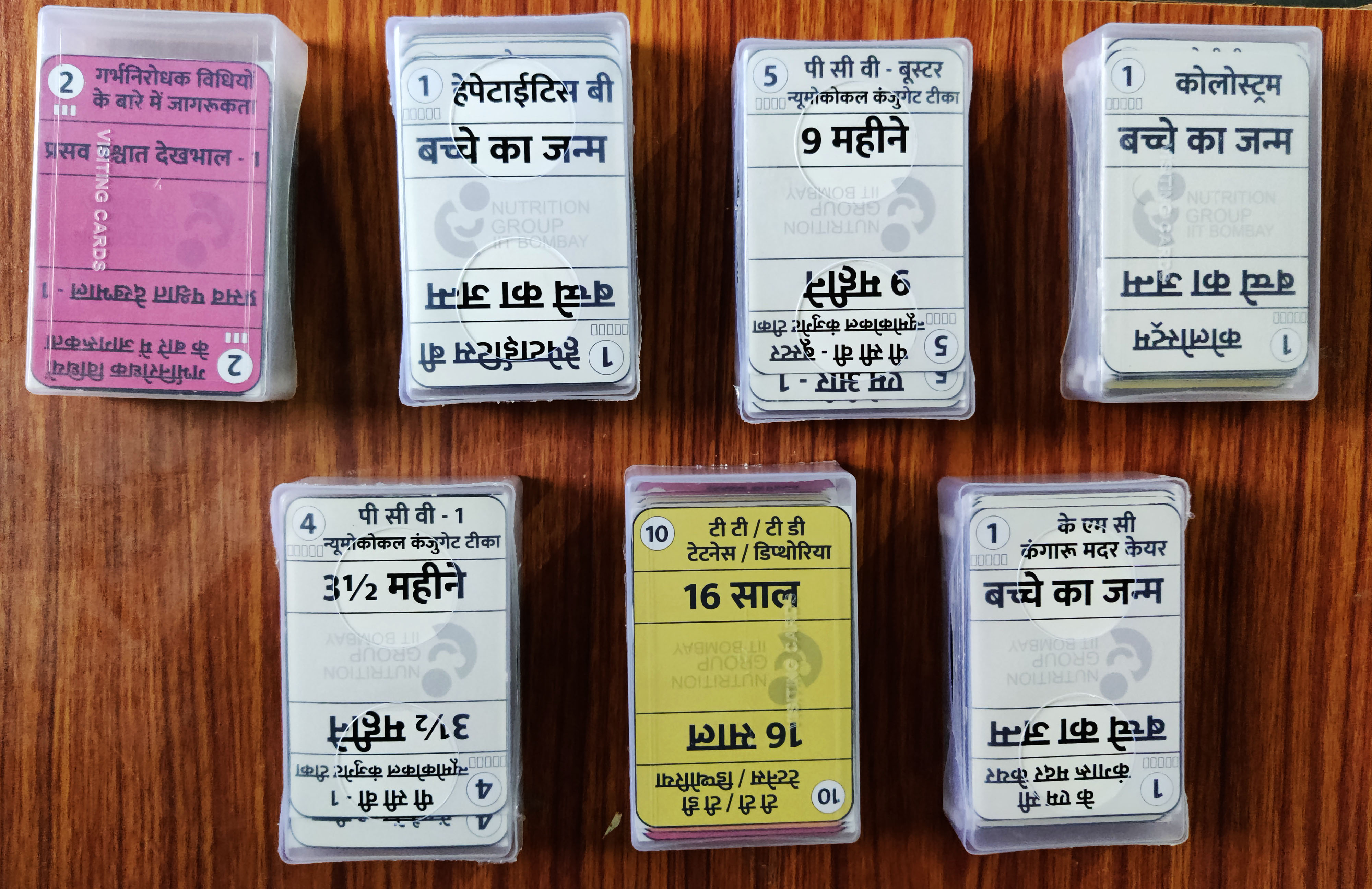}
  \caption{Physical card Decks}
  \label{fig: Physical_Cards}
\end{figure}

\subsection{Participants}

The sample size was calculated using G-Power \cite{Erdfelder2007}. T-test with independent means was chosen as the statistical test (d=0.5, 1-\(\beta\)=0.95, and \(\alpha\)=0.05). The required sample or number of participants for each group was calculated to be 88. Considering attrition, 95 were recruited for each group. However, post-attrition, 90 participants from each group were retained till the end of the study, totaling 270 participants across three groups.

\begin{table*}
  \begin{center}    
  
  \begin{tabular}{|l|c|c|c|c|c|c|}
\hline
\multicolumn{7}{|c|}{\textbf{A. Participants' Demography}}\\
\hline
 & \multicolumn{2}{c|}{\textbf{Intervention Group-1}} & \multicolumn{2}{c|}{\textbf{Intervention Group-2}} & \multicolumn{2}{c|}{\textbf{Intervention Group-3}} \\
 & \multicolumn{2}{c|}{\textbf {Digital Card Game}} & \multicolumn{2}{c|}{\textbf {Physical Card Game}} & \multicolumn{2}{c|}{\textbf {Classroom}} \\
 & \multicolumn{2}{c|}{} & \multicolumn{2}{c|}{} & \multicolumn{2}{c|}{\textbf {In-Person training}} \\
 & \multicolumn{2}{c|}{(n=90)} & \multicolumn{2}{c|}{(n=90)} & \multicolumn{2}{c|}{(n=90)} \\
 \hline
 \textbf {Parameters} & ASHAs(\%) & AWWs(\%) & ASHAs(\%) & AWWs(\%)) & ASHAs(\%) & AWWs(\%) \\
  & (n=45) & (n=45) & (n=45) & (n=45) & (n=45) & (n=45) \\
 \hline
 \textbf {Age group (years)} &&&&&&  \\
 \hline
Less than 30               & 1 (2.22\%)    & 0 (0\%)        & 1 (2.22\%)   & 1 (2.22\%)    & 2 (4.44\%)   & 2 (4.44\%)   \\ \hline
30-40                      & 22 (48.89\%)  & 19 (42.22\%)   & 22 (48.89\%) & 20 (44.44\%)  & 23 (51.11\%) & 19 (42.22\%) \\ \hline
40-50                      & 21 (46.67\%)  & 25 (55.56\%)   & 20 (44.44\%) & 24 (53.33\%)  & 20 (44.44\%) & 24 (53.33\%) \\ \hline
No information             & 1 (2.22\%)    & 1 (2.22\%)     & 2 (4.44\%)   & 0 (0\%)       & 0 (0\%)      & 0 (0\%)      \\ \hline

 \textbf {Education} &&&&&&  \\
 \textbf {(schooling grade)} &&&&&&  \\
 \hline

Below  $8^{th}$                  & 3 (6.67\%)    & 2 (4.44\%)     & 1 (2.27\%)   & 2 (4.44\%)    & 1 (2.22\%)   & 2 (4.44\%)   \\ \hline
$8^{th}-10^{th}$                      & 9 (20\%)      & 11 (24.44\%)   & 9 (20.45\%)  & 14 (31.11\%)  & 10 (22.22\%) & 12 (26.67\%) \\ \hline
$10^{th}-12^{th}$                      & 29 (64.44\%)  & 28 (62.22\%)   & 29 (65.91\%) & 23 (51.11\%)  & 29 (64.44\%) & 27 (60\%)    \\ \hline
Graduate and above         & 4 (8.89\%)    & 4 (8.89\%)     & 4 (9.09\%)   & 5 (11.11\%)   & 5 (11.11\%)  & 4 (8.89\%)   \\ \hline
No information             & 0 (0\%)       & 0 (0\%)        & 1 (2.27\%)   & 1 (2.22\%)    & 0 (0\%)      & 0 (0\%)      \\ \hline

 \hline
 \textbf {Experience as}&&&&&&  \\
\textbf {CHWs (years)} &&&&&&  \\
\hline
0-5                        & 2 (4.44\%)    & 2 (4.44\%)     & 4 (8.89\%)   & 3 (6.67\%)    & 4 (8.89\%)   & 2 (4.35\%)   \\ \hline
5-10                       & 18 (40\%)     & 17 (37.78\%)   & 16 (35.56\%) & 16 (35.56\%)  & 19 (42.22\%) & 17 (36.96\%) \\ \hline
10-15                      & 18 (40\%)     & 18 (40\%)      & 18 (40\%)    & 19 (42.22\%)  & 17 (37.78\%) & 18 (39.13\%) \\ \hline
Above 15                   & 7 (15.56\%)   & 8 (17.78\%)    & 7 (15.56\%)  & 6 (13.33\%)   & 4 (8.89\%)   & 8 (17.39\%)  \\ \hline
No information             & 0 (0\%)       & 0 (0\%)        & 0 (0\%)      & 1 (2.22\%)    & 1 (2.22\%)   & 1 (2.17\%)   \\ \hline
 
\hline
& \multicolumn{2}{c|}{\textbf{}} & \multicolumn{2}{c|}{\textbf{}} & \multicolumn{2}{c|}{\textbf{}} \\

\textbf{Experimental} & \multicolumn{2}{c|}{\textbf{Intervention Group-1}} & \multicolumn{2}{c|}{\textbf{Intervention Group-2}} & \multicolumn{2}{c|}{\textbf{Intervention Group-3}} \\

\textbf{Study Results} & \multicolumn{2}{c|}{\textbf{Test Scores}} & \multicolumn{2}{c|}{\textbf{Test Scores}} & \multicolumn{2}{c|}{\textbf{Test Scores}} \\

 \hline
 ~ & Mean & Median & Mean & Median & Mean & Median \\
 Test phase & (SD) & (min-max)  & (SD) & (min-max)  & (SD) & (min-max)   \\
 \hline
  \multicolumn{1}{|l|}{Pre-test scores}       & \multicolumn{1}{l|}{24.85 (3.34)} & \multicolumn{1}{l|}{24.16 (18-33)} & \multicolumn{1}{l|}{24.28 (2.88)} & \multicolumn{1}{l|}{24.13 (19-30)} & \multicolumn{1}{l|}{25.02 (3.19)} & \multicolumn{1}{l|}{24.69 (19-32)} \\ \hline
\multicolumn{1}{|l|}{Post-test scores}      & \multicolumn{1}{l|}{35.62 (3.57)} & \multicolumn{1}{l|}{35.56 (28-42)} & \multicolumn{1}{l|}{31.86 (4.1)}  & \multicolumn{1}{l|}{31.48 (24-42)} & \multicolumn{1}{l|}{30.01 (4.32)} & \multicolumn{1}{l|}{29.71 (19-39)} \\ \hline
\multicolumn{1}{|l|}{Long Post-test scores} & \multicolumn{1}{l|}{31.75 (4.69)} & \multicolumn{1}{l|}{32.23 (22-41)} & \multicolumn{1}{l|}{30.62 (4.1)}  & \multicolumn{1}{l|}{30.82 (21-40)} & \multicolumn{1}{l|}{26.46 (3.54)} & \multicolumn{1}{l|}{26.75 (19-34)} \\ 
  (After three weeks)&&&&&&\\
  \hline

  \hline
\multicolumn{7}{|c|}{\textbf{B. Between-group comparisons} of total scores according to the three points of assessment }\\
\hline
 & \multicolumn{2}{c|}{\textbf{Pre-test}} & \multicolumn{2}{c|}{\textbf{Post-test}} & \multicolumn{2}{c|}{\textbf{Long-term Post-test}} \\
 
\hline
IG-1 X IG-2 X IG-3          & \multicolumn{2}{c|}{$F_{2,45}=0.84,p=.44$} & \multicolumn{2}{c|}{$F_{2,45}=22.96,p<.00001$} & \multicolumn{2}{c|}{$F_{2,45}=20.62,p<.00001$} \\
\hline

\hline
\multicolumn{7}{|c|}{\textbf{C. Within-group comparisons} of total scores according to the three points of assessment }\\
\hline
\textbf{Paired t-test results} & \multicolumn{2}{c|}{\textbf{Intervention Group-1}} & \multicolumn{2}{c|}{\textbf{Intervention Group-2}} & \multicolumn{2}{c|}{\textbf{Intervention Group-3}} \\
 
 \textbf{Points of Assessment} & \multicolumn{2}{c|}{(p-value)} & \multicolumn{2}{c|}{(p-value)} & \multicolumn{2}{c|}{(p-value)}  \\
\hline
Pre-test X Post-test            & \multicolumn{2}{c|}{0} & \multicolumn{2}{c|}{0} & \multicolumn{2}{c|}{1 X 10$^{-7}$} \\
\hline
Pre-test X Long-term            & \multicolumn{2}{c|}{2 X 10$^{-10}$} & \multicolumn{2}{c|}{0} & \multicolumn{2}{c|}{0.04} \\
\qquad \qquad \quad Post-test                       & \multicolumn{2}{c|}{} & \multicolumn{2}{c|}{} & \multicolumn{2}{c|}{} \\
\hline
Post-test X Long-term           & \multicolumn{2}{c|}{6 X 10$^{-5}$} & \multicolumn{2}{c|}{0.12} & \multicolumn{2}{c|}{6 X 10$^{-5}$} \\
\qquad \qquad \quad Post-test                       & \multicolumn{2}{c|}{} & \multicolumn{2}{c|}{(Not Significant at $p < 0.05$)} & \multicolumn{2}{c|}{} \\
\hline

\hline
\multicolumn{7}{|c|}{\textbf{D. Between-group comparisons} of total scores according to the three points of assessment }\\
\hline
\multicolumn{3}{|c|}{\textbf{Paired t-test results (All groups combined)}} & \multicolumn{4}{c|}{\textbf{Intervention Group-1}} \\
 
\multicolumn{3}{|c|}{\textbf{Point of Assessment}}  & \multicolumn{4}{c|}{(p-value)}  \\
\hline
\multicolumn{3}{|c|}{Pre-test X Post-test}                      & \multicolumn{4}{c|}{0} \\
\hline
\multicolumn{3}{|c|}{Pre-test X Long-term Post-test}            & \multicolumn{4}{c|}{2 X 10$^{-20}$} \\
\hline
\multicolumn{3}{|c|}{Post-test X Long-term Post-test}           & \multicolumn{4}{c|}{7 X 10$^{-7}$} \\
\hline

 \end{tabular}
 \end{center}
 \caption{A. Participants' Demography and B. \& C. Study Results}
  \label{tab:ParticipantsDemography}
 \end{table*}

 \subsection{Ethical consideration}
The Institute Review Board, Indian Institute of Technology Bombay, India \cite{IRB_IITB}, approved the ethical conduct of the study (Approval No: IITB-IRB/2022/051), which adheres to the guidelines of the Declaration of Helsinki \cite{DOH2013}. All participants were verbally informed about the objective and procedure of the study. Written consent with a signature was obtained every time a survey was conducted with the participants and included in the questionnaire. The results of the surveys and study findings were provided to the CHW supervisors so that they could understand the overall knowledge gained and retention of the CHWs. However, the names and other identifiers of the participants were masked.

 \subsection{Evaluation}
The knowledge of CHWs is assessed through questionnaire surveys in three points. Initially, a baseline survey is conducted to check the prior knowledge. Then, as per the allotted groups, they either get a deck of cards or install the app on their phones. Then, the CHWs either play the card game on their smartphones (IG1) or the physical card game (IG2) or attend regular in-person classroom training (IG3). Then, another assessment is conducted to check the immediate knowledge gained after the refresher training. Then, CHWs are encouraged to play and attend classes according to their allotted groups for the next three weeks. Previous researchers suggest two or more weeks for retention tests \cite{Haynie1994,Nungester1982}. After that, a long-term post-test survey is conducted to check the knowledge retention of the CHWs.

The questionnaire starts with consent to participate in the evaluation study by filling in demographic details and putting in a signature. It contains 45 multiple-choice, single-correct questions of equal weightage and is shuffled for each assessment point. CHWs are usually given 30 minutes to fill out the questionnaire. Researchers and volunteers check the answers against a solved questionnaire and calculate the score for each participant. They are then analyzed to check for trends in knowledge gain and retention.

 \begin{figure*} 
\begin{tabular}{c c}
\includegraphics[width=0.325\textwidth]{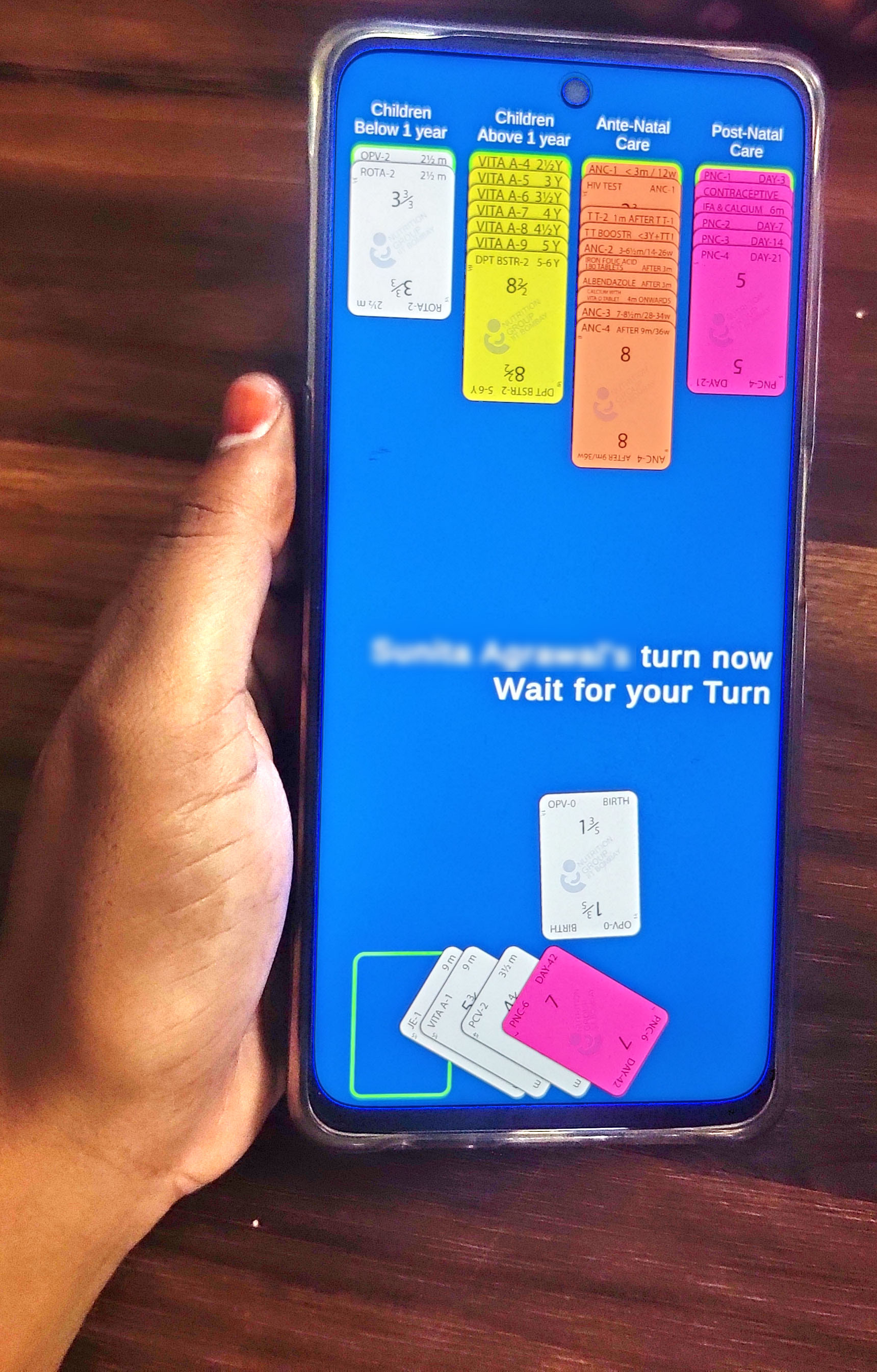} &
\includegraphics[width=0.675\textwidth]{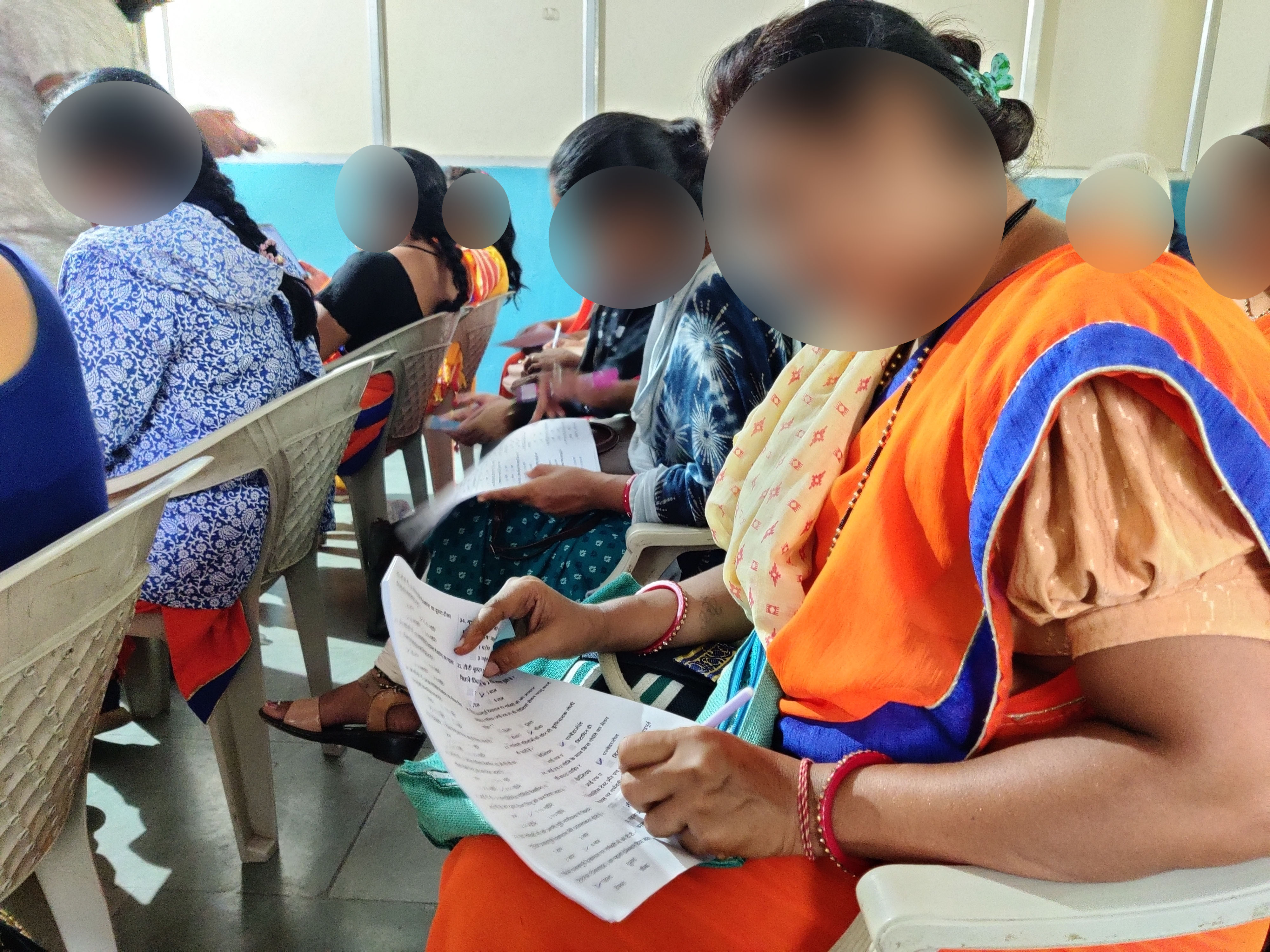} 
  
\end{tabular}
\caption{ Left: Photo of CHW holding her smartphone during gameplay session; Right: CHWs filling questionnaire survey form}
\label{fig:questionnaire_phone}
\end{figure*}

 All group scores were checked for normality before conducting parametric tests. Paired t-tests were conducted within groups and between groups. Repeated Measure ANOVA and multiple One-Way ANOVA were performed to check if there was an overall difference in scores within groups and between groups and with points of assessment.

\section{Findings}

 \begin{figure*} 
\includegraphics[width=\textwidth]{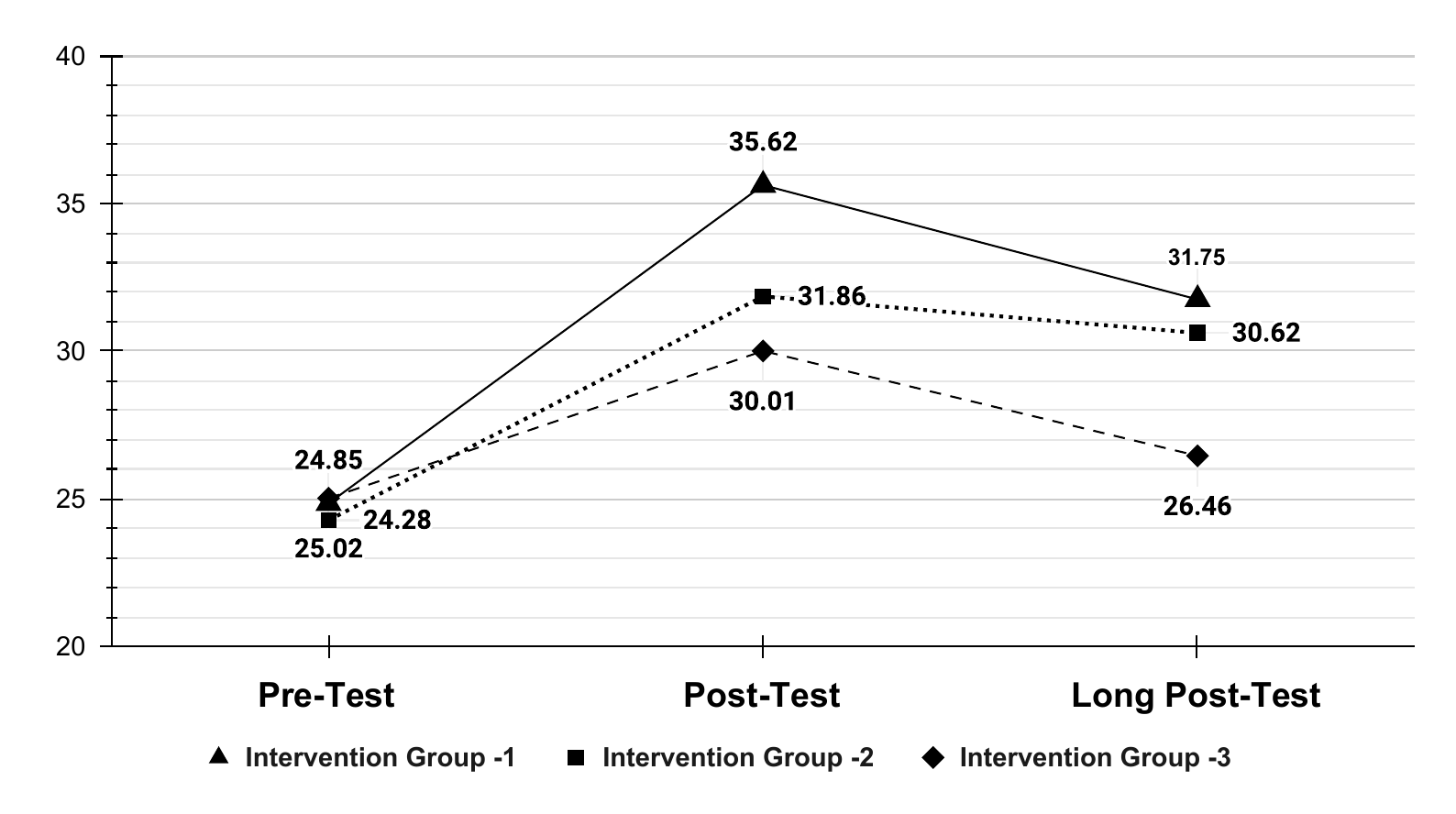}
\caption{Line chart showing trends of change of mean value in the pre-test, post-test, and long post-test}
\label{fig:knowledge_graph}
\end{figure*}

\subsubsection{Between group (Table \ref{tab:ParticipantsDemography} B): }

\begin{description}
    \item [Pre-test scores:] No significant differences were found between groups ($F=0.84,p=.44$) at $p<.05$. We found that the baseline or prior knowledge was not significantly different across the three groups.
    \item [Post-test scores:] Significant differences were found between groups ($F=22.96,p<.00001$) at $p<.05$. Tukey's HSD test shows pairwise significant difference ($Q_{.05} = 3.35$) between IG1 and IG2 (Q=6.36 (p=.00004)) \& IG3 (Q=9.39 (p<.000001)) and no significance between IG2 and IG3 (Q=3.03 (p=.085)). IG1, or smartphone app intervention group, scored exceptionally better than IG2, or physical card game, and IG3, or in-person training. 
    \item [Long-term Post-test scores:] Significant difference between groups ($F=20.62,p<.00001$) at $p<.05$. Tukey's HSD test shows pairwise significant difference ($Q_{.05} = 3.35$) between IG3 and IG1 (Q=8.62 (p<.000001)) \& IG2 (Q=6.78 (p=.00001)) and no significance between IG1 and IG2 (Q=1.84 (p=.4)).
\end{description}

\subsubsection{Within group (Table \ref{tab:ParticipantsDemography} C): }
A significant difference was observed between all three groups across all assessment points except IG2 while comparing post-test and long-term post-test (p=.12, p>.05)

\subsubsection{Between group (Table \ref{tab:ParticipantsDemography} D): }
Pre-test scores were significantly lower compared to post-tests and long-term post-tests. A significant difference was observed between all three groups across all assessment points. A decline in knowledge retention across the groups confirms the pattern of depleting knowledge over time. Comparing digital and physical card game outcomes, it's interesting to note that the immediate knowledge gained post-intervention in smartphone mode or digital one (IG1) is significantly higher compared to physical card play mode (IG2). However, the long-term post-test scores showed lower knowledge retention in all three groups. IG2 had the highest retention rate, and the loss of knowledge was minimal compared to the baseline or pretest scores. After three weeks, the retained knowledge or long-term post-test scores were similar or not significantly different between digital and physical play modes or IG1 and IG2. The knowledge graph of the three groups (Fig: \ref{fig:knowledge_graph}) also reflects the findings.

\section{Discussion}

In this study, we compared two modes of refresher training for CHWs: traditional in-person refresher training and training through games (RQ1). Also, we compared if there is a difference in knowledge acquisition between physical and digital card play (RQ2). Also, we compared if there is a difference in knowledge retention between physical and digital card play (RQ3). The study evaluates the efficacy of three refresher training methods. It provides objective evidence to curriculum designers and decision-makers on implementing the method and the expected short- and long-term deployment outcome.

After the study, we interviewed some CHWs for feedback. Some CHWs expressed their preference for printed study materials over the phone app. They said that the printed materials improve their knowledge and act as a vetted brochure to show to mothers during counseling sessions or home visits. Some said the printed brochure lets them relax and study independently, at their own pace. Some reported distractions when playing smartphone games for refresher training compared to reading class notes or printed brochures.

\section{Limitations}

This is a field-based study and was conducted in government institutions like government hospitals and training centres. Some CHWs might feel intimidated by their supervisor's presence while giving feedback on the play experience. Often, one smartphone was shared by multiple CHWs. Some felt uncomfortable using someone else's devices due to the possessiveness and affordances that users grow using their own devices. Conducting a true experiment is impossible in this case because it does not fulfill the condition of a true experiment \cite{Harris2006}. So, we conducted a quasi-experimental study. The drawback of such a study lies in lower internal and higher external validity, as the effectiveness is judged through the outcome \cite{Harris2006}.

\section{Novelty}

As far as we know, this research attempted to explore the effects of combining two CHW cadres, ASHAs and AWWs, as a team to play physical and digital card games for their refresher training for the first time.

\section{Conclusion}

The study results show that the digital play mode enhanced knowledge better than the physical mode. However, the digital mode has shown a marginal improvement in long-term knowledge retention compared to the physical mode. The latter finding is more important, as knowing the long-term effects of an intervention is required for curriculum development, implementation, policy recommendations, and future decision-making. It is important to note that the gamified training method is not intended to replace the initial in-person training of the CHWs but to act as a voluntary refresher training for the CHWs to help them revise the child immunization schedule and other services.

\subsection{Future Works}

Further studies need to be conducted, considering other variables that might affect the intrinsic and extrinsic motivation of playing the game. Longitudinal studies need to be undertaken to check improved learning outcomes, with more arms for checking variations in learning effectiveness and a larger sample size for a more balanced and better representation of CHWs in India.

\begin{credits}
\subsubsection{\ackname} This study was jointly funded by SERB, FICCI, and UNICEF India. We thank all the Community Healthcare Workers (Accredited Social Health Activists and Anganwadi Workers) and their supervisors who actively participated in this study. We also thank the anonymous reviewers for providing feedback.

\subsubsection{\discintname}
The authors have no conflicts of interest.

\end{credits}

\bibliographystyle{splncs04}

\bibliography{template}

@techreport{Gopalan2012,
    title = {{Addressing maternal healthcare through demand side financial incentives: experience of Janani Suraksha Yojana program in India}},
    institution = {The World Bank, Washington, DC 20433, USA},
    year = {2012},
    author = {Gopalan, Saji S and Varatharajan, Durairaj},
    url = {http://www.biomedcentral.com/1472-6963/12/319}
}

@article{Kizhatil2019,
    title = {{Assessment of immunization coverage and associated factors among children in Paravur Taluk of Ernakulam district, Kerala}},
    year = {2019},
    journal = {International Journal Of Community Medicine And Public Health},
    author = {Kizhatil, Anuradha and ., Reshma and Hariharan, Harsha Chollankil and John, Alexander and Thomas, Ann Mary and Padmanabhan, Gopalakrishnan},
    number = {8},
    pages = {3594},
    volume = {6},
    doi = {10.18203/2394-6040.ijcmph20193494},
    issn = {2394-6032}
}

@article{Farah2015,
    title = {{Assessment of ‘Accredited Social Health Activists’—A National Community Health Volunteer Scheme in Karnataka State, India}},
    journal = {J Health Popul Nutr},
    year = {2015},
    author = {Farah, N. Fathima and Mohan, Raju and Kiruba, S. Varadharajan and Aditi, Krishnamurthy and S, R. Ananthkumar and Prem, K. Mony}
}

@inproceedings{Javaid2017,
    title = {{Bridging the knowledge gaps in lady health visitors through video based learning tool}},
    year = {2017},
    booktitle = {ACM International Conference Proceeding Series},
    author = {Javaid, Maham and Fatima, Beenish and Batool, Amna},
    volume = {Part F1320},
    doi = {10.1145/3136560.3136603}
}

@techreport{Haynie1994,
    title = {{Effects of Multiple-Choice and Short-Answer Tests on Delayed Retention Learning}},    
    institution = {Department of Occupational Education, North Carolina State University, Raleigh, NC},
    year = {1994},
    booktitle = {Journal of Technology Education},
    author = {Haynie, W J},
    number = {1},
    pages = {1994--2026},
    volume = {6}
}

@article{Bag2017,
    title = {{Evaluation of mother and child protection card entries in a rural area of West Bengal}},
    year = {2017},
    journal = {International Journal Of Community Medicine And Public Health},
    author = {Bag, Subhadip and Datta, Mousumi},
    number = {7},
    pages = {2604},
    volume = {4},
    doi = {10.18203/2394-6040.ijcmph20172867},
    issn = {2394-6032}
}

@misc{Bashingwa2021,
    title = {{Examining the reach and exposure of a mobile phone-based training programme for frontline health workers (ASHAs) in 13 states across India}},
    year = {2021},
    booktitle = {BMJ Global Health},
    author = {Bashingwa, Jean Juste Harrisson and {et.al.}},
    month = {8},
    volume = {6},
    publisher = {BMJ Publishing Group},
    doi = {10.1136/bmjgh-2021-005299},
    issn = {20597908}
}

@article{Burger2022,
    title = {{Facilitating behavioral change: A comparative assessment of ASHA efficacy in rural Bihar}},
    year = {2022},
    journal = {PLOS Global Public Health},
    author = {Burger, Oskar and {et.al.}},
    number = {8},
    month = {8},
    pages = {e0000756},
    volume = {2},
    publisher = {Public Library of Science (PLoS)},
    doi = {10.1371/journal.pgph.0000756}
}

@article{Yadav2019a,
    title = {{FeedPal: Understanding opportunities for chatbots in breastfeeding education of women in India}},
    year = {2019},
    journal = {Proceedings of the ACM on Human-Computer Interaction},
    author = {Yadav, Deepika and {et.al.}},
    number = {CSCW},
    volume = {3},
    doi = {10.1145/3359272},
    issn = {25730142}
}

@inproceedings{Erdfelder2007,
    title = {{G*Power 3: A flexible statistical power analysis program for the social, behavioral, and biomedical sciences}},
    year = {2007},
    booktitle = {Behavior Research Methods},
    author = {Faul, Franz and Erdfelder, Edgar and Lang, Albert Georg and Buchner, Axel},
    volume = {39},
    doi = {10.3758/BF03193146},
    issn = {1554351X}
}

@article{Ward2020,
    title = {{Impact Of Mhealth Interventions for Reproductive, Maternal, Newborn and Child Health and Nutrition at Scale: Bbc Media Action and The Ananya Program in Bihar, India}},
    year = {2020},
    journal = {Journal of Global Health},
    author = {Ward, Victoria C. and {et.al.}},
    number = {2},
    volume = {10},
    doi = {10.7189/jogh.10.021005},
    issn = {20472986}
}

@article{Kasi2021,
    title = {{Indian Academy of Pediatrics (IAP) Advisory Committee on Vaccines and Immunization Practices (ACVIP): Recommended Immunization Schedule (2020–21) and Update on Immunization for Children Aged 0 Through 18 Years}},
    year = {2021},
    journal = {Indian Pediatrics},
    author = {Kasi, Srinivas G. and {et.al.}},
    number = {1},
    volume = {58},
    doi = {10.1007/s13312-021-2096-7},
    issn = {09747559}
}

@misc{IRB_IITB,
    title = {{Institutional Review Board, IIT Bombay, India}},
    year = {2020},
    author = {{IIT Bombay}},
    month = {5},
    url = {https://rnd.iitb.ac.in/institute_review_board}
}

@article{Yadav2019,
    title = {{LEAP: Scaffolding collaborative learning of community health workers in India}},
    year = {2019},
    journal = {Proceedings of the ACM on Human-Computer Interaction},
    author = {Yadav, Deepika and Bhandari, Anushka and Singh, Pushpendra},
    number = {CSCW},
    volume = {3},
    doi = {10.1145/3359271},
    issn = {25730142}
}

@article{Kumar2015,
    title = {{Mobile phones for maternal health in rural India}},
    year = {2015},
    journal = {Conference on Human Factors in Computing Systems - Proceedings},
    author = {Kumar, Neha and Anderson, Richard},
    pages = {427--436},
    volume = {2015-April},
    isbn = {9781450331456},
    doi = {10.1145/2702123.2702258}
}

@article{Ramachandran2010,
    title = {{Mobile-izing health workers in rural India}},
    year = {2010},
    journal = {Conference on Human Factors in Computing Systems - Proceedings},
    author = {Ramachandran, Divya and Canny, John and Das, Prabhu Dutta and Cutrell, Edward},
    pages = {1889--1898},
    volume = {3},
    isbn = {9781605589299},
    doi = {10.1145/1753326.1753610}
}

@article{NFHS4IndiaFactsheet,
    title = {{National Family Health Survey-4 India Fact Sheet}},
    year = {2016},
    journal = {Indian Institute Population Sciences and Ministry of Health and Family Welfare},
    author = {{Indian Institute Population Sciences and Ministry of Health and Family Welfare}},
    url = {http://rchiips.org/NFHS/NFHS-4Reports/India.pdf}
}

@article{NFHS5IndiaFactsheet,
    title = {{National Family Health Survey-5 India Fact Sheet}},
    year = {2021},
    journal = {Indian Institute Population Sciences and Ministry of Health and Family Welfare},
    author = {{Indian Institute Population Sciences and Ministry of Health and Family Welfare}},
    url = {http://rchiips.org/nfhs/factsheet_NFHS-5.shtml}
}

@inproceedings{Majhi2022,
    title = {{Physical and Augmented Reality based Playful Activities for Refresher Training of ASHA Workers in India}},
    year = {2022},
    booktitle = {Conference on Human Factors in Computing Systems - Proceedings},
    author = {Majhi, Arka and Agnihotri, Satish and Mondal, Aparajita},
    doi = {10.1145/3516492.3558788}
}

@inproceedings{Kumar2015a,
    title = {{Projecting Health: Community-Led Video Education for Maternal Health}},
    year = {2015},
    booktitle = {ACM International Conference Proceeding Series},
    author = {Kumar, Neha and {et.al.}},
    volume = {15},
    doi = {10.1145/2737856.2738023}
}

@inproceedings{Majhi2021,
    title = {{Refresher Training through Quiz App for capacity building of Community Healthcare Workers or Anganwadi Workers in India}},
    year = {2021},
    booktitle = {5th Asian CHI Symposium 2021},
    author = {Majhi, Arka and Joshi, Anirudha and Agnihotri, Satish B. and Mondal, Aparajita},
    doi = {10.1145/3429360.3468186}
}

@inproceedings{Yadav2017,
    title = {{Sangoshthi: Empowering community health workers through peer learning in rural India}},
    year = {2017},
    booktitle = {26th International World Wide Web Conference, WWW 2017},
    author = {Yadav, Deepika and {et.al.}},
    doi = {10.1145/3038912.3052624}
}

@article{Tulaskar2020,
    title = {{Study of Instructional Illustrations on ICTs: Considering persona of low-literate users from India}},
    year = {2020},
    journal = {ACM International Conference Proceeding Series},
    author = {Tulaskar, Rucha},
    pages = {53--56},
    isbn = {9781450387682},
    doi = {10.1145/3391203.3391217}
}

@inproceedings{Shah2017,
    title = {{Tackling child malnutrition: An innovative approach for training health workers using ICT: A pilot study}},
    year = {2017},
    booktitle = {IEEE Region 10 Humanitarian Technology Conference 2016, R10-HTC 2016 - Proceedings},
    author = {Shah, Mithilesh P. and Kamble, Pawan A. and Agnihotri, Satish B.},
    isbn = {9781509041770},
    doi = {10.1109/R10-HTC.2016.7906811}
}

@techreport{Nungester1982,
    title = {{Testing Versus Review: Effects on Retention}},
    institution={The American College},
    year = {1982},
    booktitle = {Journal of Educational Psychology},
    author = {Nungester, Ronald J and Duchastel, Philippe C},
    number = {1},
    pages = {18--22},
    volume = {74}
}

@article{Verdezonto2021,
    title = {{The Invisible Work of Maintenance in Community Health: Challenges and Opportunities for Digital Health to Support Frontline Health Workers in Karnataka, South India}},
    year = {2021},
    journal = {Proceedings of the ACM on Human-Computer Interaction},
    author = {Verdezoto, Nervo and Bagalkot, Naveen and Akbar, Syeda Zainab and Sharma, Swati and MacKintosh, Nicola and Harrington, Deirdre and Griffiths, Paula},
    number = {CSCW1},
    month = {4},
    volume = {5},
    publisher = {Association for Computing Machinery},
    doi = {10.1145/3449165},
    issn = {25730142}
}

@inproceedings{Kumar2013,
    title = {{The mobile media actor-network in urban India}},
    year = {2013},
    booktitle = {Conference on Human Factors in Computing Systems - Proceedings},
    author = {Kumar, Neha and Rangaswamy, Nimmi},
    pages = {1989--1998},
    isbn = {9781450318990},
    doi = {10.1145/2470654.2466263}
}

@article{Harris2006,
    title = {{The use and interpretation of quasi-experimental studies in medical informatics}},
    year = {2006},
    journal = {Journal of the American Medical Informatics Association},
    author = {Harris, Anthony D. and {et.al.}},
    number = {1},
    month = {1},
    pages = {16--23},
    volume = {13},
    doi = {10.1197/jamia.M1749},
    issn = {10675027},
    pmid = {16221933}
}

@techreport{TRAI_Nov_23,
    title = {{TRAI Press Release No.03/2024}},
    institution={TRAI},
    year = {2024},
    author = {{Telecom Regulatory Authority of India}},
    url = {https://www.trai.gov.in/sites/default/files/PR_No.03of2024_0.pdf}
}

@misc{DOH2013,
    title = {{World Medical Association Declaration of Helsinki: Ethical principles for medical research involving human subjects}},
    year = {2013},
    booktitle = {JAMA},
    author = {{WMA}},
    number = {20},
    volume = {310},
    doi = {10.1001/jama.2013.281053},
    issn = {15383598}
}

\end{document}